\documentclass[12pt,preprint]{aastex}

\newcommand{\msunyr}{$M$\mbox{$_{\odot}$}{\rm yr}$^{-1}$}

\newcommand{\kms}{\,km~s$^{-1}$}
\newcommand{\EW}{$W_{\lambda}$}

\newcommand{\Feii}             {Fe~{\sc ii}}
\newcommand{\Tiii}             {Ti~{\sc ii}}
\usepackage[bf,small]{caption}

\slugcomment{To appear in Ap.\ J., .}

\shorttitle{The axi-symmetric wind of IRC +10420} \shortauthors{Davies \&
Oudmaijer}

\begin{document}


\title{Integral-Field Spectroscopy of the Post Red Supergiant
IRC~+10420: evidence for an axi-symmetric wind}


\author{Ben Davies\altaffilmark{1,2}, Ren\'{e}
  D. Oudmaijer\altaffilmark{2} and Kailash C. Sahu\altaffilmark{3} 
}
\affil{$^{1}$Center for Imaging Science, Rochester Institute of Technology,
  54 Lomb Memorial Drive, \\Rochester, NY 14623, USA \\
$^{2}$School of Physics \& Astronomy, University of Leeds, Woodhouse
  Lane, \\ Leeds LS2 9JT, UK \\
$^{3}$Space Telescope Science Institute, 3700 San Martin Drive,
  Baltimore, MD 21218, USA}

\begin{abstract}
We present {\it NAOMI/OASIS} adaptive-optics assisted integral-field
spectroscopy of the transitional massive hypergiant IRC +10420, an
extreme mass-losing star apparently in the process of evolving from a
Red Supergiant toward the Wolf-Rayet phase. To investigate the
present-day mass-loss geometry of the star, we study the appearance of
the line-emission from the inner wind as viewed when reflected off the
surrounding nebula. We find that, contrary to previous work, there is
strong evidence for wind axi-symmetry, based on the equivalent-width
and velocity variations of H$\alpha$ and Fe {\sc ii} $\lambda$6516. We
attribute this behaviour to the appearance of the complex
line-profiles when viewed from different angles. We also speculate
that the Ti {\sc ii} emission originates in the outer nebula in a
region analogous to the Strontium Filament of $\eta$ Carinae, based on
the morphology of the line-emission. Finally, we suggest that the
present-day axisymmetric wind of IRC~+10420, combined with its
continued blueward evolution, is evidence that the star is evolving
toward the B[e] supergiant phase.
\end{abstract}





\section{Introduction}
\object{IRC +10420} ($19^{h} 26^{m} 48.0^{s} ~+11\degr 21' 16.7''$,
J2000) is an extremely luminous star which inhabits the so-called
`yellow-void' on the HR diagram between the Red and Blue Supergiants
\citep{dJ98}. There is considerable evidence that the star is evolving
rapidly away from the RSG phase towards the Luminous Blue Variable
(LBV) or Wolf-Rayet (WR) phases (see below). As such this object
represents a potential link between the key post-MS mass-losing
stages, and is therefore considered extremely important for the study
of massive stellar evolution.

The star is surrounded by a dusty circumstellar nebula, responsible
for its large IR-excess. This nebula is the result of an extreme
mass-losing episode in the RSG phase, when the mass-loss reached $\ga$
10$^{-4}$\msunyr\ \citep[][ O96 hereafter]{Oudmaijer96}. It is also an
OH maser source, a phenomenon usually identified with much cooler
stars \citep{Giguere76}. This maser is thought to be a relic of an
earlier M supergiant phase, from which the central star must have
rapidly evolved on a timescale of $\la10^{5}$ years in order for the
maser source to still survive \citep{Mutel79}. Indeed, the star has
been observed to gradually increase in temperature from $\sim$6000\,K
to $\sim$9200\,K in the last 30 years \citep[O96,][]{Klochkova02}.

The material the star has recently ejected may be the precursor to a
LBV/WR nebula. These nebulae are often axisymmetric, but it is unclear
how these nebulae are formed. Hydrodynamical studies have shown that
\hbox{axisymmetric} morphologies can arise from either an axisymmetric
wind in the LBV/WR phase, or a spherically-symmetric wind ploughing
into a slower axisymmetric wind ejected in the RSG phase
\citep{Frank95,D-O02}. As wind-axisymmetry may be linked to rotation,
which itself plays an important role in the evolution of a massive
star \citep[see e.g.\ review of][]{M-M00}, and as IRC +10420 appears to
be somewhere between the LBV/WR and RSG phases, determining the true
geometry of the star's present-day wind may provide insight into the
formation mechanism of the bipolar nebulae of massive stars. Further,
it may also provide clues as to the role of rotation in its subsequent
evolution and the connection to other classes of massive star.

The geometry of IRC +10420's wind has been the topic of much
discussion over the last $\sim$20 years, yet a common concensus
remains elusive due to the weight of apparently contradictory
evidence. The OH maser emission was spatially-resolved by
\citet{Diamond83}, who suggested the masing emission originated in an
equatorial outflow seen almost edge-on. In the mid 1980s, the star
began to show H$\alpha$ emission \citep{Irvine86}, high-resolution
observations of which showed it to be doubly-peaked \citep{Jones93} --
reminiscent of classical Be stars which are commonly thought to have
outflowing disks \citep[][]{P-R03}. Additionally, Jones et al.\ argued
that while the star has considerable IR-excess, there is only modest
extinction towards the central star. This, they concluded, was due to
the circumstellar material being located in an inclined disk, with the
star itself relatively unobscurred. Also, their mid-IR imaging was
elongated at a PA of $\sim$150\degr, which they suggested was the PA
of the disk.

However, from near infra-red spectra, \citet{Oudmaijer94} argued that
the H {\sc i} emission was inconsistent with a circumstellar disk,
suggesting instead that the ionized material was located in a bipolar
outflow oriented close to the line-of-sight. \citet{Humphreys97}
argued for bipolarity of the {\it outer} nebula, and that the
morphology was analogous to that of the dusty homunculus nebula around
$\eta$~Car oriented almost pole-on. They suggested the extended
emission to the SW was due to the nearer lobe, whilst the receding
lobe on the far-side to the NE was obscurred by equatorial material.

The confusion was apparently complete when \citet{Humphreys02}, using
{\it HST/STIS} long-slit spectroscopy, utilized the star's reflection
nebula as a tool with which to view the H$\alpha$ emission from
different angles. With the slit aligned along the long-axis of the
nebula, they showed that the velocities of the two peaks in their
H$\alpha$ profile were `surprisingly uniform'. This they argued was
inconsistent with a circumstellar disk or a bipolar outflow, and
concluded that the star showed {\it no evidence of axisymmetry} in the
present-day wind.

However, Humphreys et al. only observed two slit positions, each
roughly parallel to the long-axis of the nebula and separated by
0.5\arcsec. In this paper, their data is improved on with
spatially-resolved spectroscopy across the {\it whole} of the inner
nebula, with integral-field spectroscopy. We will show that, contrary
to the conclusions of Humphreys et al., the emission-lines {\it are}
variable when viewed from different angles. It is argued that this is
due to the non-isotropic nature of the star's wind, and that the
line-emission is formed in an axi-symmetric structure in the inner
wind.



We begin in Sect.\ \ref{sec:ifuobs} with a description of the
observations, data-reduction steps and analysis techniques. The
results of the data analysis are presented and discussed in Sect.\
\ref{sec:ifuresults}, and summarized in Sect.\ \ref{sec:ifuconc}.


\section{Observations \& Data-Reduction}\label{sec:ifuobs}
\subsection{Observations}
Data was taken during 3-5 September 2005 at the WHT on La Palma, using
{\it OASIS} in {\it Tiger} mode \citep{Bacon95} in conjuction with the
{\it NAOMI} adaptive optics system \citep{Myers03}. Observing
conditions were good, with a small amount of extinction ($\sim$ few
tenths of 1 mag) due to Saharan dust. Seeing was 0.6-1.0\arcsec\
throughout our programme.

The 22mm enlarger was used in conjunction with the HR667 filter, which
gives a plate-scale of 0.97\AA/pixel in the wavelength range 6490 -
6840\AA\ and a lenslet size of 0.26\arcsec.

In addition to the usual calibrations of biases, dome-flats and neon
arcs, the following extra calibration exposures must be taken for {\it
OASIS} -- undispersed micropupil frames, which record the positions of
the micropupils focussed by the lenslet arrays onto the CCD; and
continuum frames, which record the positions of the spectral `ridges'
on the CCD. These calibrations were done by illuminating the lenslets
with a tungsten bulb.

For the science frames the observational strategy employed was that of
exposures with increasing integration times, to gradually increase the
signal-to-noise ratio (SNR) in the outer parts of the field-of-view
whilst obtaining unsaturated data at the centre of the
image. Exposures were dithered by $\sim$0.1\arcsec to identify and
reject bad-pixels on the CCD. Overall, 6 hours of science exposures
were obtained in 17 separate exposures.

In addition, a PSF standard star, HIP95166, was observed in order to
accurately measure the PSF at the time of observations. The standard
is separated from IRC +10420 by 1.3\degr, and has brightess $V =$10.4
compared with $V=$11.1 for IRC +10420. 

In practice, even the longest exposure times of 30min only just
saturated the H$\alpha$ line in the central couple of pixels. However,
the MITTL3 CCD is very susceptible to cosmic rays, and long exposure
images were so decimated by cosmic hits that the spectral mask
creation algorithm failed in the outer regions where the SNR is
lower. Subsequently, some data had to be discarded. Therefore, for
future observations using this setup, integration times of no longer
than $\sim$15min are recommended.

\subsection{Data reduction}
All data-reduction was done with the instrument's custom-written
software {\sc xoasis}\footnote{{\tt
http://www.cfht.hawaii.edu/Instruments/Spectroscopy/OASIS/Reduc/}}, and
following the steps outlined in the software documentation. Initially,
a median-averaged bias frame was subtracted from all calibration and
science frames. A spectral 'mask' was created using the micropupil and
continuum frames, which recorded the loci of each micropupil spectrum
on the detector. This mask was used to extract each spectrum of a
given exposure, which were arranged in a datacube with the spatial
coordinates of the corresponding lenslet.

Each spectrum in a single-exposure datacube was flat-fielded with the
corresponding spectrum in the master continuum-lamp
datacube. Wavelength calibration was done by fitting a 2nd-degree
polynomial to the five identified neon lines in the observed
wavelength range. When identifying and rejecting cosmic rays, it was
found that the {\sc xoasis} algorithm was unable to deal with the high
number of hits without attacking the H$\alpha$ line centre. Instead,
cosmics were rejected at the mosaicing stage.

To mosaic the 17 individual frames, each were first recentered about
the brightness-centre of each exposure. The frames were weighted
according to the mean number of counts per frame, and resampled onto a
square spatial sampling grid using bicubic interpolation. The median
spectrum was then found at each spatial point on the grid, thus
eliminating cosmic ray hits. As the observations were dithered by
$\sim$0.1\arcsec, we chose to spatially sample the individual frames
onto a master grid of 0.13\arcsec per pixel, i.e. half the size of a
lenslet.

\subsection{Data analysis} \label{sec:anal}
The continuum of each spectrum was fitted with a 2nd-degree
polynomial, using only those regions deemed to be featureless from
the high-SNR integrated over the whole field. Thus, a datacube of the
continuum at each spatial pixel was created.

The discrete features of the spectral lines -- see e.g.\
\citet{Oudmaijer98} -- are poorly resolved in these data, so spatial
variations in line velocities were measured by fitting gaussian
profiles to the lines in each spectrum of the datacube. In order to
test the uniformity of the instrumental spectral resolution, this
analysis was applied to the Neon arc frames. It was found that the
spectral resolution was $60 \pm 5$\kms\ over the whole chip.

Synthetic narrow-band images were created by first summing the spectra
in the datacube over the width of the line. The corresponding
continuum was found by summing over the same spectral region in the
`continuum-cube' (the continuum-fit to the datacube, see above). The
`pure' line-image was then found by subtracting the `continuum' image
from the `continuum+line' image.

For PSF characterization and subtraction, the spatial PSF of the
observations was assumed to be that of the standard star. As no
spectral information was required of this object, the spectra of the
standard were aggressively cleaned for cosmics and then median
averaged in the spectral direction to produce a 2-D image. This was
then resampled onto the same spatial grid as the target object. The
FWHM of this image was found to be $\sim$0.5\arcsec. This corresponds
to a Strehl ratio of $\sim$0.1, and an improvement of a factor of 2 on
the seeing conditions. This is consistent with the expected
performance of {\it NAOMI} at optical wavelengths. Qualitative
PSF-subtraction was done by iteratively scaling the flux of the PSF
standard with the flux of the science target, until cancellation
effects at the centre of the image were minimized.

\begin{figure}[t]
  \centering
  \includegraphics[width=12cm,bb=30 0 580 470,clip]{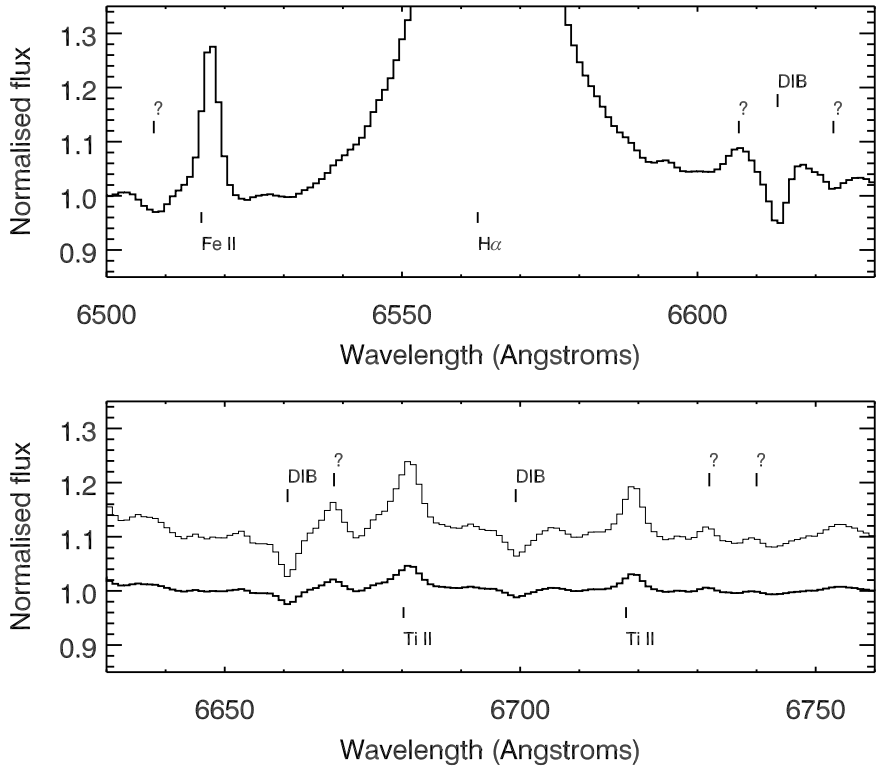}
  \caption{The spectrum of IRC + 10420, integrated over the whole field
  of view. The lighter line in the bottom panel shows the spectrum
  magnified to highlight the weaker features. }
  \label{fig:spec}
\end{figure}

\section{Results \& Discussion} \label{sec:ifuresults}
In this section, the results of the datacube analysis are presented
and discussed. All spatial images shown are centred on 19$^{h}$
26$^{m}$ 48.0$^{s}$~+11\degr\ 21$\arcmin$ 16.7$\arcsec$ (J2000), with
north at the top and east to the left. To aid the interpretation of
the data, the images are overlayed with a contour map of a {\it HST}
F451M image taken from the archive, with a spatial resolution of
$\sim$0.1\arcsec. As the image was taken in the blue part of the
spectrum, the reflection nebula is clearly seen.

\subsection{Mean spectrum of IRC +10420}
\begin {table}
    \caption[Spectral features of the central region.]{Spectral
    features of the central region, marked as $e$ in
    Fig.\,\ref{fig:spregmap}. Column (1): the measured line-centres
    are in the heliocentric frame; (2): line identifications if any
    (see notes below); (3): equivalent width, with error $\pm$5\%
    determined from repeatability of measurements when different
    continuum points either side of the line were used; (4):
    heliocentric radial velocity, with error taken to be $\pm$5\kms
    $\sim$0.1 pixels; (5): FWHM of a gaussian fit to the line, with
    error $\pm$5\% determined from repeatability of
    measurements. \label{tab:ifuspec} }
  \begin {center}
    \begin {tabular}{llrcc}
\hline \hline
$\lambda_{\rm hel}$ (\AA)~~~&Ident.&$W_{\lambda}$ (\AA)&$v_{\rm hel}$
      &FWHM  \\
 & & & (\kms) & (\kms) \\
\hline
6508.08 & $^{1}$ & +0.16 & - & 198 \\
6517.39&Fe {\sc ii}&-1.0&+62&151 \\
6564.57&H$\alpha$&-50.2&+80&372 \\
6594.90&[Ti {\sc ii}] ?&-0.04&+90&109 \\
6607.07&$^2$&-0.28&-&193 \\
6613.53&DIB&+0.23&+1&117 \\
6618.64&$^3$&-0.15&-&199 \\
6627.51&$^3$&-0.12&-&195 \\
6660.59&DIB&+0.09&-2&108 \\
6668.08&$^4$&-0.08&-&167 \\
6680.82&Ti {\sc ii}$^5$&-0.30&+25&249 \\
6699.61&DIB&+0.05&+17&147 \\
6719.02&Ti {\sc ii}&-0.14&+50&171 \\
\hline
    \end{tabular}
  \end {center}
\footnotesize $^{1}$Unidentified line, Ne {\sc i}? \\ 
$^{2}$Resolved
in O98 as four discrete components, and in \citet{Chentsov99} as
two. Identified by the latter as a blend of Sc {\sc ii} and Ti {\sc
ii}. \\ 
$^{3}$Unidentified lines, may be split components of the same
line -- see Sect. \ref{sec:un}. \\ 
$^{4}$Unidentified line, possibly a blend
of Fe {\sc i} and/or [Ni {\sc ii}]? \\ 
$^{5}$Possible unresolved blend of Ti
{\sc ii} and He {\sc i} $\lambda$6678
\\ 
\label{tab:ifuspec}
\end {table}

\begin{figure}[t]
  \centering
  \includegraphics[width=12cm,clip,bb=35 0 568 438]{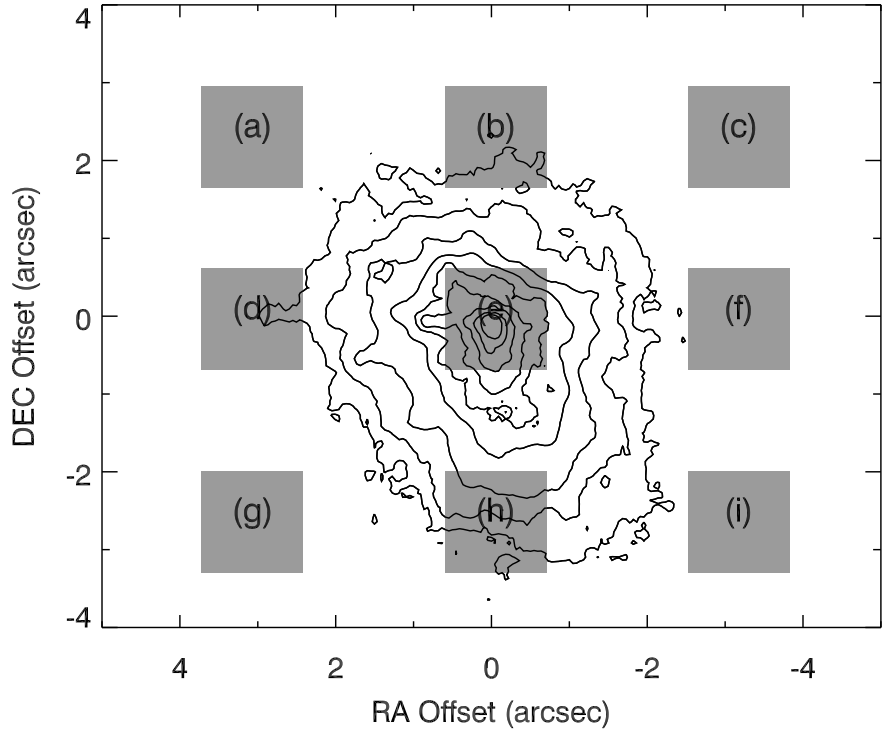}
  \caption{Definition of the 9 spatial regions over which the
  spectra were integrated. }
  \label{fig:spregmap}
\end{figure}

The spectrum over the full wavelength range observed and integrated
over the whole image is shown in Fig.\,\ref{fig:spec}. The individual
spectra were flux-weighted before adding, and continuum-flattened
later. In addition to the strong H$\alpha$ line, the permitted lines
of Fe {\sc ii} $\lambda$6516 and Ti {\sc ii} $\lambda\lambda$6680,
6717 are identified. Also, there are several unidentified
features. The emission bump on the blue side of the DIB at $\lambda
6614$ is identified by \citet{Chentsov99} as a blend of Sc {\sc ii}
and Ti {\sc ii}. The other unidentified spectral features are not
present in Oudmaijer's 1994 spectrum. The properties of observed
spectral features of the central region (region ({\it e}) in
Fig.\,\ref{fig:spregmap}) are listed in Table \ref{tab:ifuspec}.

The spectral features identified in Table 1 do not all exhibit the
same behaviour across the field. Below, we emphasize this by showing
the residual spectrum between the central star and the surrounding
nebula. This is followed by a discussion of the behaviour of
individual lines. As the signal-to-noise (S/N) of the lines falls-off
with distance from the star, this analysis is limited to those with
the highest S/N, in order to assess the spatial variations over as
large a field as possible. These features are the H$\alpha$, Fe {\sc
ii} and Ti {\sc ii} emission lines, as well as the unidentified blend
at $\sim6610$\AA ~and the diffuse interstellar bands (DIBs).

\begin{figure}[t]
  \centering
  \includegraphics[width=12cm,bb=0 0 613 485]{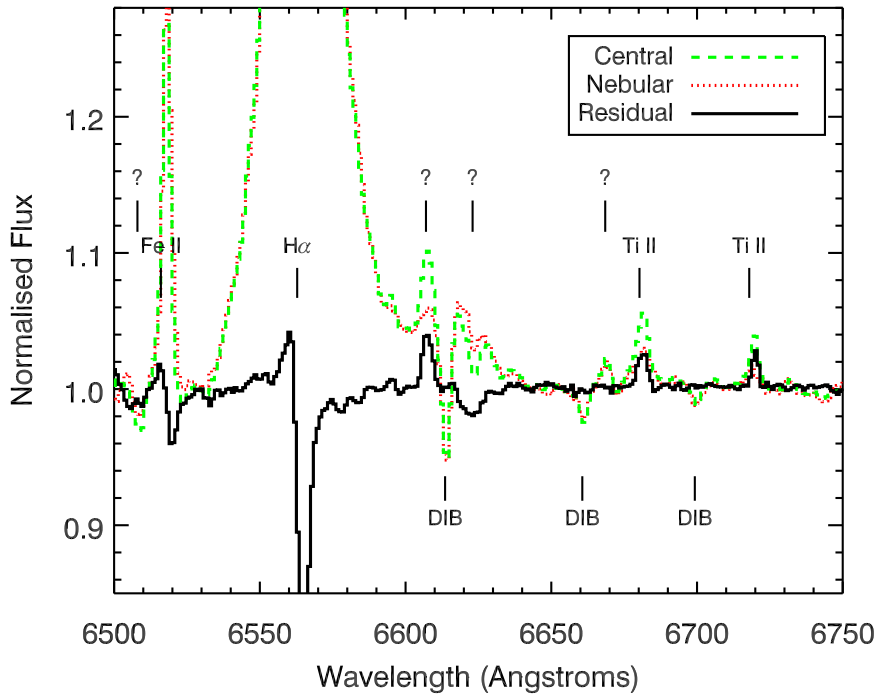}
  \caption{The residual spectrum (solid line) after the normalised
  spectrum of the central region (dashed line) is divided by the
  normalised mean nebular spectrum (dotted line). See text for a
  discussion of the identified features.}
  \label{fig:subspec}
\end{figure}

\subsubsection{Residual spectrum of central object} \label{sec:resid}
In order to find the differences between the spectra of the central
object and the surrounding nebula, we divide the field up into nine
spatial regions: a central (star+nebula) region, surrounded by eight
nebula regions (illustrated in Fig.\ \ref{fig:spregmap}). The spectra
of the outer regions were added together and normalised by the mean
continuum spectrum of the same regions, to ensure that the total
`nebular' spectrum is weighted towards the higher S/N regions. The
normalised spectrum of the central region (region ($e$) in
Fig.\,\ref{fig:spregmap}) was then divided by the total nebular
spectrum. 

The resulting spectrum is shown in Fig.\,\ref{fig:subspec}. While the
DIBs are seen to vanish, the Ti {\sc ii} lines remain, as does the
unidentified line at $\sim$6607\AA. The features around the H$\alpha$
and Fe {\sc ii} lines result from the ratio of lines of different
width and velocities. In the following sections, the lines are
individually discussed in more detail.

\subsection{Diffuse interstellar bands} \label{sec:dibs}
There are three DIBs in the spectral region covered -- one in the
red-wing of H$\alpha$ at 6614\AA\ and two in the region of the Ti {\sc
ii} lines at 6661\AA\ and 6700\AA. Figure \ref{fig:subspec} shows that
these lines all cancel-out perfectly in the residual spectrum between
the star and the surrounding nebula. Analysis of the DIB at 6700\AA,
the only DIB which appears not to be blended with other spectral
features, shows that the equivalent width of the line is constant over
the field to within the errors. However, the S/N in the outer regions
is low, and for this reason it is not illustrated here.

The cancellation of the DIBs in the residual spectrum is an important
result. These features are indicative of the interstellar extinction
toward IRC~+10420 \citep{Oudmaijer98}, and appear to be equally strong
towards the central star and the outer nebula. This supports the
conclusions of \citet{Jones93} who found that, whilst the star has
significant IR-excess, there is only a relatively small amount of {\it
circumstellar} extinction in the line-of-sight to the central star. In
addition, this result strongly suggests that any variations observed
in other spectral features are real.

\subsection{The H$\alpha$ and Fe {\sc ii} $\lambda$6516 emission}
\begin{figure}[t]
  \centering
  \includegraphics[width=8.cm,bb=10 10 535 400]{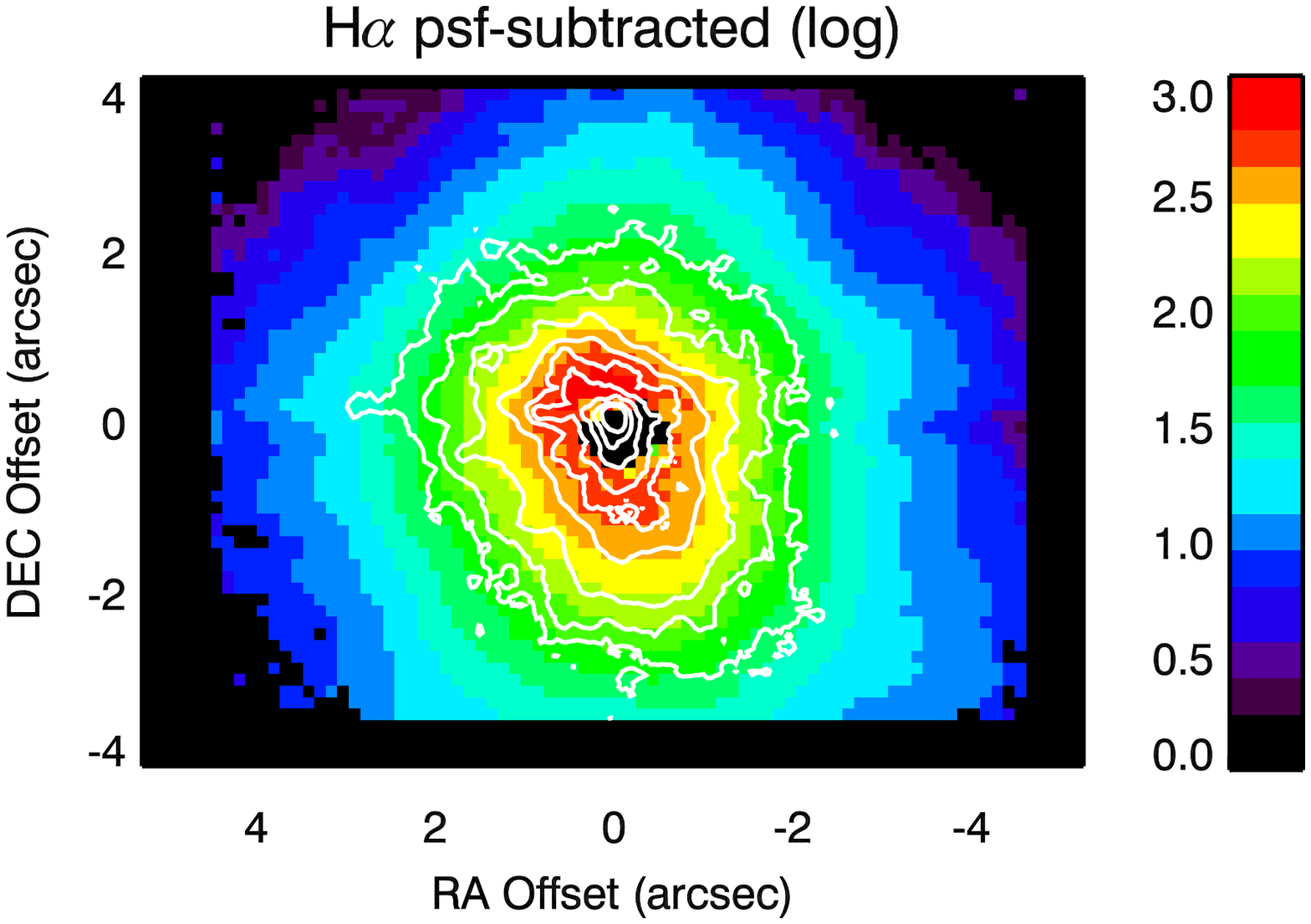}
  \includegraphics[width=8.cm,bb=10 10 535 400]{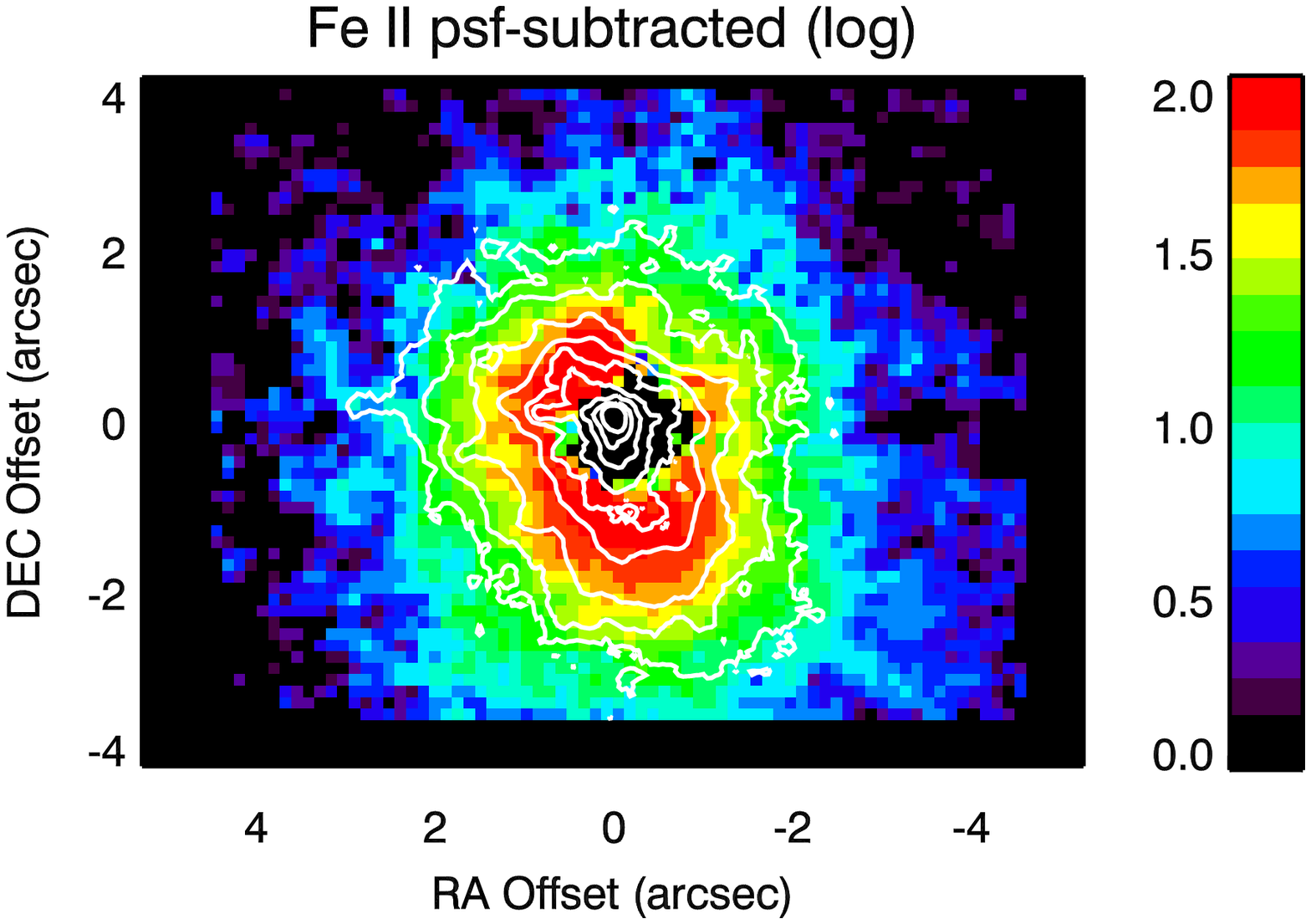}
  \caption{Continuum-subtracted, PSF-subtracted image of IRC +10420 in
  the light of H$\alpha$ ({\it left}) and Fe {\sc ii} ({\it
  right}). The log of the image has been taken, and is scaled in units
  of $\sigma$ above the sky background as measured from the outer
  regions. The image is overlayed with a contour map of the $B$-band
  {\it HST} image. The sharp decrease in intensity west of the central
  star in the H$\alpha$ image is an artifact caused by a mis-match in
  the diffraction spikes of the PSF standard and IRC +10420. Both
  images show the line-emission tracing the contours of the reflection
  nebula.}
  \label{fig:hapsf}
\end{figure}

The H$\alpha$ and Fe {\sc ii} emission, as traced by the
continuum-subtracted, PSF-subtracted images, closely follow the
morphology of the reflection nebula as shown in the continuum $B$-band
{\it HST} image (see Fig.\,\ref{fig:hapsf}). A star with $T_{\rm eff}
\sim 9200$K is unlikely to emit enough flux shortward of 912\AA\ to
ionize the circumstellar material out to these distances ($\sim$0.1pc
if $D$ = 6\,kpc is assumed). It is therefore more likely that the
line-emission forms close to the star, and the light seen from the
nebula is reflected. Further, it can be argued that the line-emission
is unresolved by the nebula: from modelling of speckle interferometry
observations, \citet{Blocker99} found an inner dust radius of
70$R_{\star}$, meaning that the star has an apparent radius of
$\sim$50\arcmin\ when viewed from the inner dust wall. If all
line-emission forms within a few stellar radii, it is reasonable to
assume that the line-formation regions are spatially-unresolved when
viewed at the distance of the circumstellar dust.

\subsubsection{Variations in line-strength}
With this in mind, we analyse the equivalent-width (\EW) variations of
the two lines. We note that the spectral resolution of this data
($\sim$60\kms) is not sufficient to resolve the double-peaked profile
in H$\alpha$ observed by \citet{Humphreys02}. However, the spectral
resolution does not affect the equivalent width \EW. Therefore any
variation in \EW\ cannot be an artifact of unresolved features but
instead must be due to a genuine change in line-profile or
line-strength across the field of view. The stark uniformity observed
in the DIBs (see Sect. \ref{sec:dibs}) would suggest that any observed
variations are real.

\begin{figure*}[t]
  \centering
  \includegraphics[width=12cm,clip,bb=10 10 535 400]{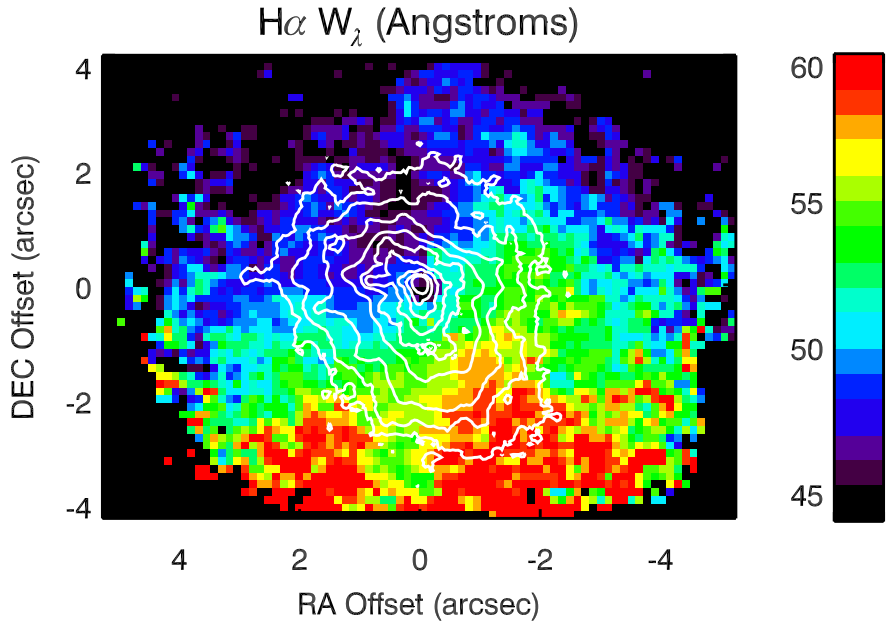}
  \caption{A map of the H$\alpha$ equivalent width ($\times ~-$1)
  across the field of view. The image is overlayed with a contour map
  of the $B$-band {\it HST} image. The figure shows that the strength
  of the H$\alpha$ emission, when viewed from the surrounding nebula,
  is strongest in the south-west and weakest in the north-east. The
  morphology is roughly axisymmetric, and perpendicular to the
  long-axis of the nebula.}
  \label{fig:haew}
\end{figure*}

A map of \EW (H$\alpha$), shown in Fig.\,\ref{fig:haew}, illustrates
that there is a marked decrease of $\sim$15\AA ~(25\%) along the long
axis of the nebula from SW to NE. We note that the same behaviour is
observed in the Fe {\sc ii} line, but at poorer signal-to-noise (not
shown). This is clear evidence of a degree of axisymmetry to
IRC~+10~420's present-day wind. 

This result is in direct contradiction to the conclusions of
\citet{Humphreys02}. In their long-slit spectra, they observed that
the velocities of the two peaks of the H$\alpha$ profile were roughly
consistent at several slit positions along the axis of the
nebula. However, their slit positions happened to be oriented
more-or-less along regions of the nebula where the H$\alpha$ \EW\ is
constant to within $\sim$5\AA. They therefore may have discounted any
observed trend in \EW\ as being too small to draw any conclusions
from.

\begin{figure}[t]
  \centering
  \includegraphics[width=8cm,bb=10 10 535 400]{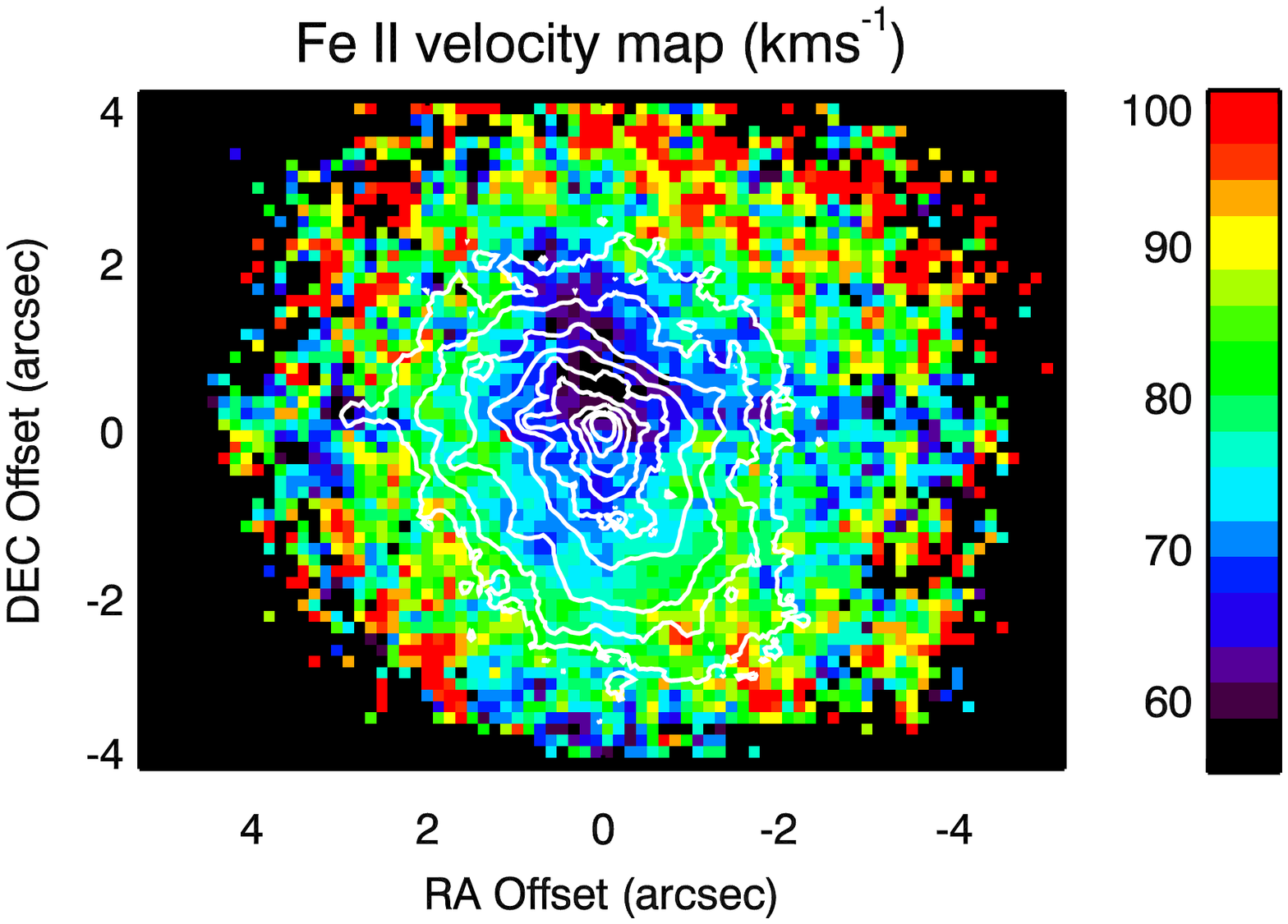}
  \includegraphics[width=8cm,bb=10 10 535 400]{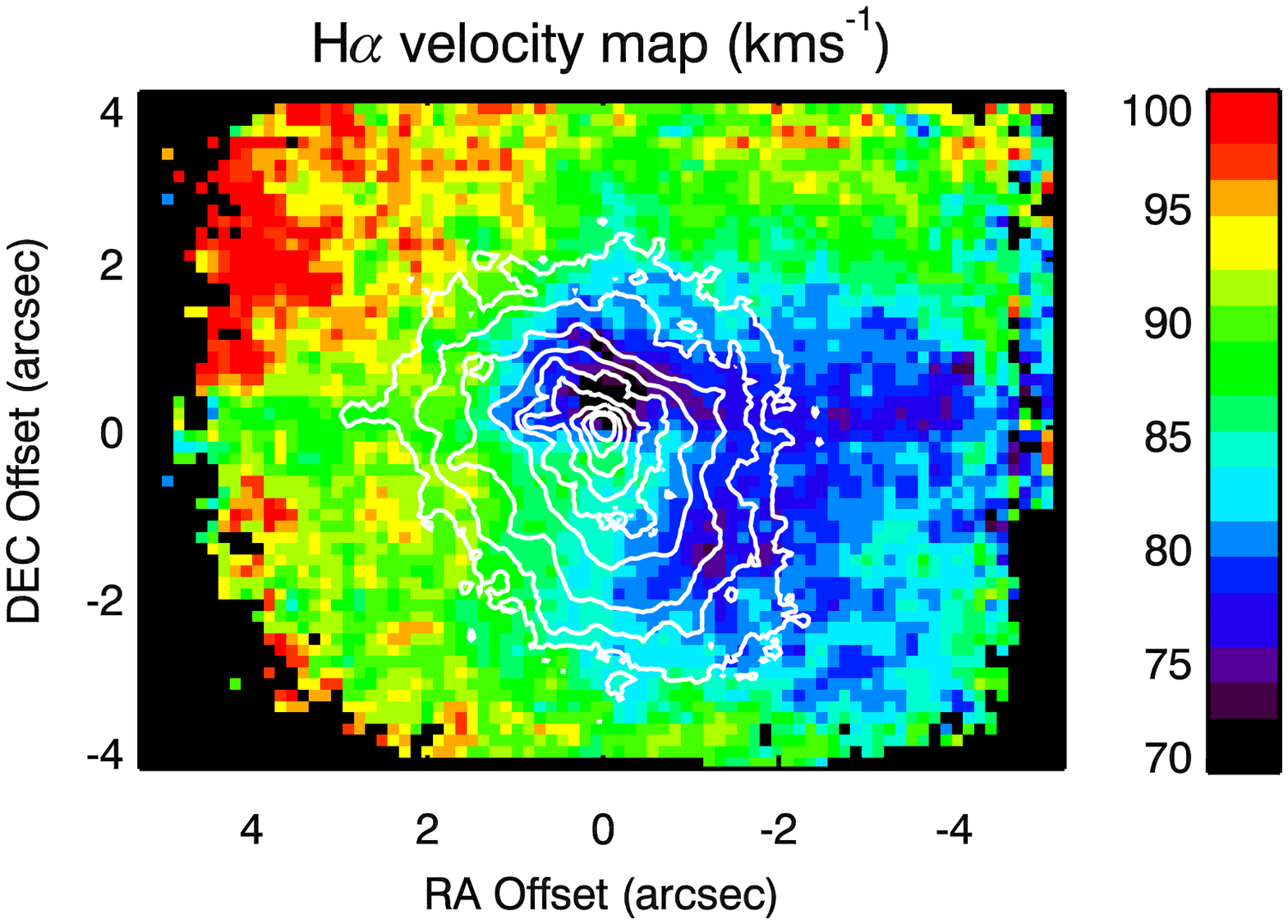}
   \caption{Velocity maps, as determined from gaussian fits to the
   emission lines. Images are heliocentric corrected, and colour bars
   to the right of each plot show the velocity scale. For clarity,
   images have been overlayed with a SNR mask such that only regions
   with a detection of 20$\sigma$ or better are shown. The two maps
   show different behaviour: the \Feii\ map shows lower velocities in
   the centre and higher velocities in the outer regions; while the
   H$\alpha$ map shows lowest velocities in the south-west and highest
   in the north-east.}
  \label{fig:velmaps}
\end{figure}

\subsubsection{Variations in velocity centroid}
As argued above, it is a valid assumption that formation regions of
the H$\alpha$ and Fe {\sc ii} lines in the stellar wind are
point-sources when viewed by the nebula. As the discrete components of
the line-profiles are unresolved in our data, at first-thought one
could assume that any shifts in the velocity centroids of the lines
may trace the nebular dynamics. If true, then the same behaviour would
be observed in both lines.

Figure \ref{fig:velmaps} shows the spatial variations of the centroids
of the Fe {\sc ii} and H$\alpha$ lines, as determined by fitting
gaussian profiles to the lines at each spatial point in the
data-cube. The kinematic morphology as traced by the Fe {\sc ii} line
appears to be slightly elongated, with the lowest velocities reached
$\sim$ 1\arcsec~ NE of the centre of the field. The velocities range
from $\sim$60\kms~ to $\sim$95\kms~ at the edge of the field.

However, contrary to the basic assumptions above, the picture is
different however in the H$\alpha$ velocity map. The kinematic
morphology of this line is somewhat axisymmetric, with regions
reaching from the centre to the SW displaying velocities of
$\sim$75\kms, and regions in the NE of $\sim$100\kms. We point out
that this is consistent with the velocities measured by
\citet{Humphreys02}, whose slit was aligned more-or-less NE-SW.

\subsubsection{Anisotropic emission?}
While asymmetry in the dynamics of the reflection nebula alone {\it
could} explain the variations in line-centroid, it {\it cannot}
explain the variations in line-strength. Further, if the H$\alpha$ and
Fe~{\sc ii} emission regions are roughly coincident (i.e.\ unresolved
by the reflection nebula), the `moving mirror' could not explain why
different velocity trends are observed in the two lines. 

If the line-emission seen by the dust is spatially unresolved, but the
line-profile had a different appearance depending on the viewing angle
to the star, this would influence the observed velocity of the line
when reflected off different regions of the nebula. The velocity maps
shown above were created by fitting gaussian profiles to the emission
lines. This is a valid assumption as the line-structure seen in
high-resolution spectra is not resolved in this data. However, if the
line-profiles are complex and anisotropic, the unresolved line-profile
may seem to be velocity-shifted when observed from different viewing
angles due to changes in the contributions from the different
components. An example of this is discribed below.

\citet{Oudmaijer98} showed in his 1994 spectrum that various Fe {\sc
ii} lines showed inverse P~Cygni profiles, which he attributed to
infalling material. In some line profiles, the absorption component
was as strong as the emission component. The separation of the
components was $\la$60\kms, and so would not necessarily be resolvable
in the data presented here. If the absorption component varied in
strength depending on the viewing angle, for example due to an
increased column-density of infalling material, it may cause the peak
of the total line-profile to shift. An increase in red-shifted
absorption would cause the peak to shift to the blue, as well as
producing an overall drop in equivalent width.

Similarly, the H$\alpha$ emission is known to be doubly-peaked, with
the blue peak typically stronger than the red
\citep[e.g.][]{Oudmaijer98,Humphreys02}. If the ratio of the two peaks,
or the depth of the central absorption, were to depend on viewing
angle, this could also cause an observed velocity trend in the
unresolved emission line, as well as a change in total line-emission.

The spectral resolution of our observations is not sufficient to
provide a detailed view of how the line-profiles change when viewed
from different angles; however from the unambiguous changes in
line-emission as probed by the \EW\ map, and the differing velocity
trends of the H$\alpha$ and Fe~{\sc ii} lines, we can say that strong
evidence exists that the wind has an axisymmetric appearance when
viewed from the surrounding reflection nebula. 

Now the angle of axisymmetry has been better constrained, at least in
terms of the H$\alpha$ emission, it would be interesting to follow-up
this study with high spectral-resolution data similar to that
presented in \citet{Humphreys02}, but with the slit position
adjusted. Such a study would allow more quantitative arguments to be
made regarding the latitudinal dependence of the star's wind
structure. Indeed, in a similar study of $\eta$~Car, \citet{Smith03}
were able to show evidence for an apparent increase in mass-flux
towards the poles.


\subsection{The Ti {\sc ii} $\lambda\lambda$6680, 6718 emission} 
\begin{figure}[t]
  \centering
  \includegraphics[width=8cm,bb=10 10 535 400]{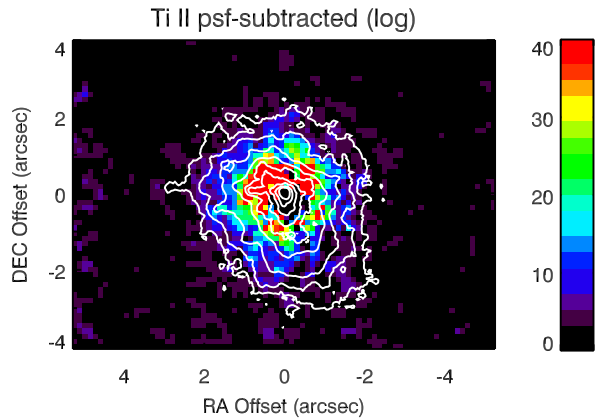}
  \caption{Continuum-subtracted, PSF-subtracted image of IRC +10420 in
  the light of Ti {\sc ii}, scaled in units of $\sigma$ above the
  sky-background. As the behaviour seen in each of the two Ti~{\sc ii}
  lines was similar, we combined the two to increase S/N. Contrary to
  similar maps of the H$\alpha$ and \Feii\ emission (Fig.\
  \ref{fig:hapsf}), the \Tiii\ emission does {\it not} trace the
  morphology of the reflection nebula. }
  \label{fig:tipsf}
\end{figure}

Although weak, it is clear that the behaviour of these lines differs
from H$\alpha$ and Fe {\sc ii} $\lambda$6516. Figure \ref{fig:tipsf}
shows the continuum-subtracted, PSF-subtracted line-image when the
light of the two lines is combined. The behaviour of each of the lines
is similar, therefore we combined the two to increase S/N.

Unlike the H$\alpha$ and Fe~{\sc ii} lines, the morphology of the
reflection nebula is not reproduced by the Ti~{\sc ii} line
images. This is not due to low S/N, as it can be seen in both the
H$\alpha$ and Fe {\sc ii} images that the S/N follows the contours of
the nebula in the {\it HST} image (Fig.\,\ref{fig:hapsf}). This would
seem to suggest that the emission zone of these lines is very
different to that of the other lines studied, as discussed below. 




\subsubsection{The origin of the \Tiii\ emission}

If the reflection-nebula paradigm is to hold, then the only reason
that the \Tiii\ emission is not reflected off the nebula is if the
line forms {\it within \rm or \it outside} the nebula, in a localised
region along our line-of-sight but a large distance from the central
star. This would also explain the radial velocities being blueshifted
by $\sim$25\kms\ compared to the systemic velocity. If correct, this
would mean that the \Tiii\ emission originates at a distance
$\sim0.05$\,pc ($D_{\star} = 5$ \,kpc) from the star.

It may be that the \Tiii\ emission comes from a region analagous to
the {\it Strontium filament} in the nebula of \object{$\eta$~Car}
\citep{Zethson01}. This is a localised region located in the
equatorial `skirt' of $\eta$~Car's homunculus nebula, which shows very
unusual spectral features of singly-ionized Sr, Ti and V amongst
others, whilst showing no detectable H or He emission and only very
weak \Feii\ emission \citep{Hartman04}. It also a source of NH$_{3}$
emission, but no H$_{2}$ emission is observed \citep{Smith06}. 

The proposed explanation for the presence of $\eta$~Car's Strontium
filament is that the emission region must be bathed in photons with $5
{\rm eV} \la E_{h \nu} \la 11{\rm eV}$ (the ionization potentials of
Sr and Ti are 5.7\,eV and 6.8\,eV respectively), but is shielded from
the intense radiation shortward of $\sim$1500\AA\ by H and \Feii\ in
the inner wind. This allows species with low ionization potentials
such as Sr~{\sc ii} and Ti~{\sc ii}, to exist over an extended region
without becoming doubly-ionized. Further, the region must be very
dense in order to produce significant Sr emission, due to its low
cosmic abundance (a factor of $\sim10^{4}$ lower than Fe)
\citep{Bautista02}.

Such a situation may exist in IRC~+10420. The \Tiii\ emission may
originate in a region located within the nebula at a distance of
$\ga$0.01\,pc ($D_{\star} = 5 \rm kpc$) along our line-of-sight,
shielded by the H and Fe in the inner wind. This would explain why the
H and Fe line-images trace the reflection nebula, while the Ti
emission, which originates from the near side of the nebula, is
essentially unseen by the rest of the nebula.

We note that this phenomenon may be less unusual in the case of
IRC~+10420. The star emits less radiation in the UV than $\eta$~Car,
and so less shielding is required for the partially-ionzed zone to
survive. Also, we have no geometry information on the other
characteristic Sr filament lines of Sr~{\sc ii} V~{\sc ii}, and
Fe~{\sc i} emission, which are present in IRC~+10420's spectrum
\citep{Oudmaijer98}. It would be interesting to compare the data
presented here with similar data which could explore the emission
regions of these lines, to confirm that they originate in the outer
nebula.

\begin{figure}
  \centering
  \includegraphics[width=10cm]{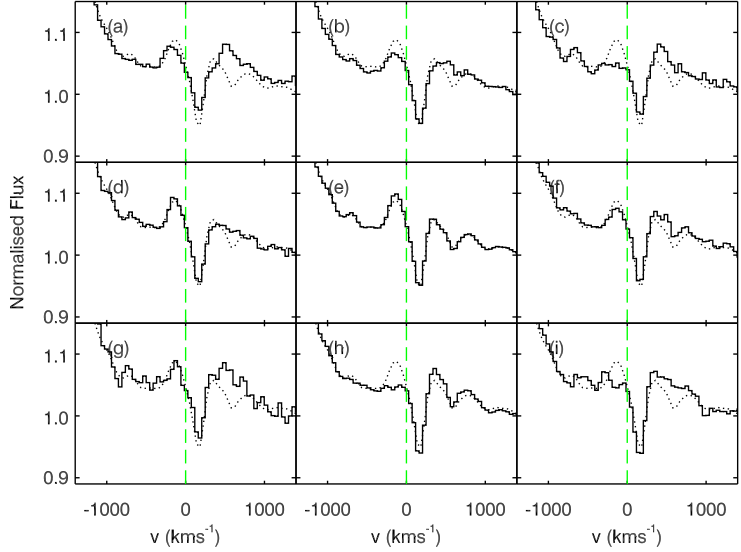}
  \caption{The unidentified features around the DIB at 6614\AA\ at the
  nine spatial regions defined in Fig.\,\ref{fig:spregmap}. Spectra
  are centred on 6610\AA.}
  \label{fig:urrprofs}
\end{figure}

\begin{figure}
  \centering
  \includegraphics[width=8cm,bb=10 50 590 530,clip]{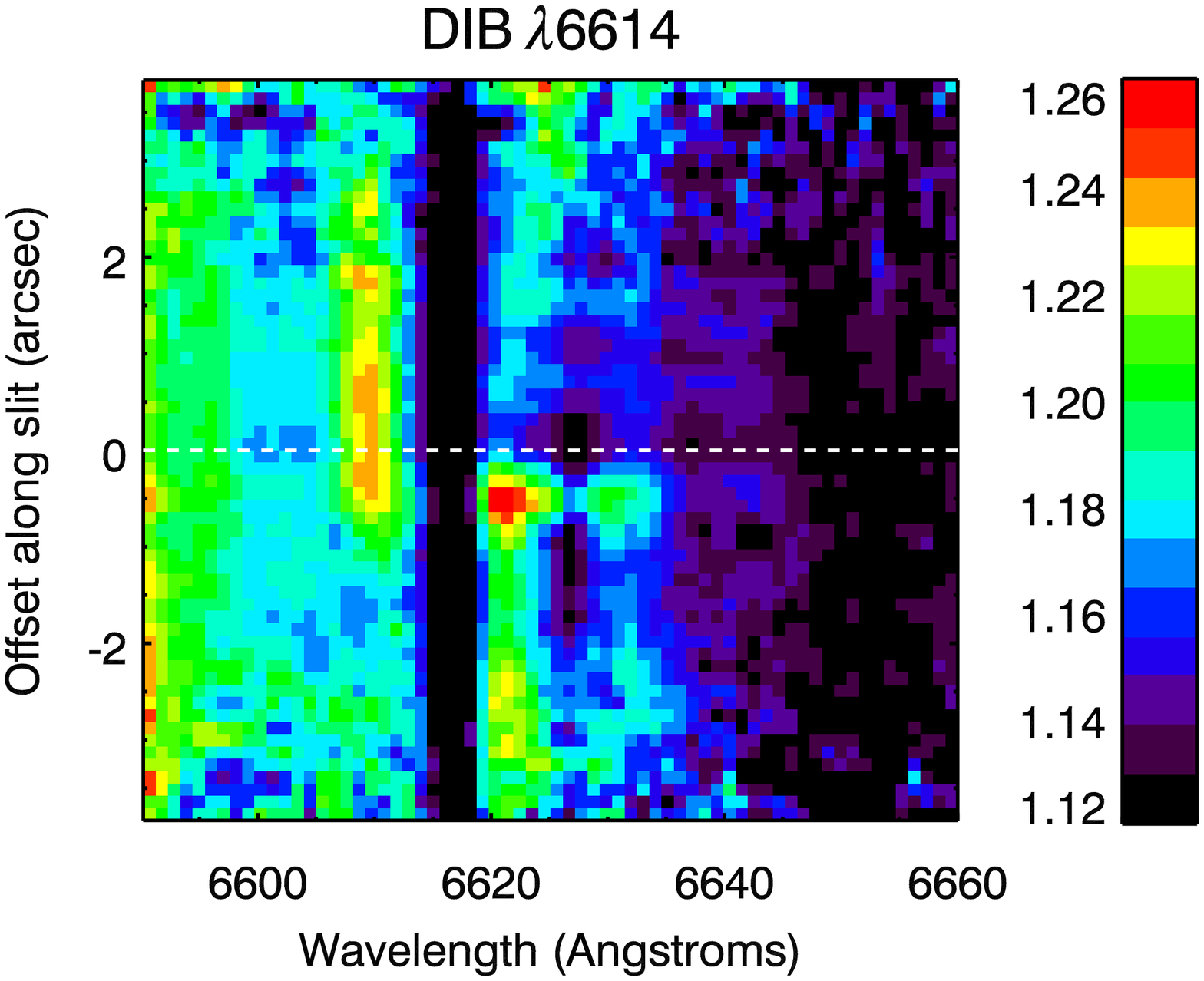}
  \includegraphics[width=8cm,bb=10 50 590 530,clip]{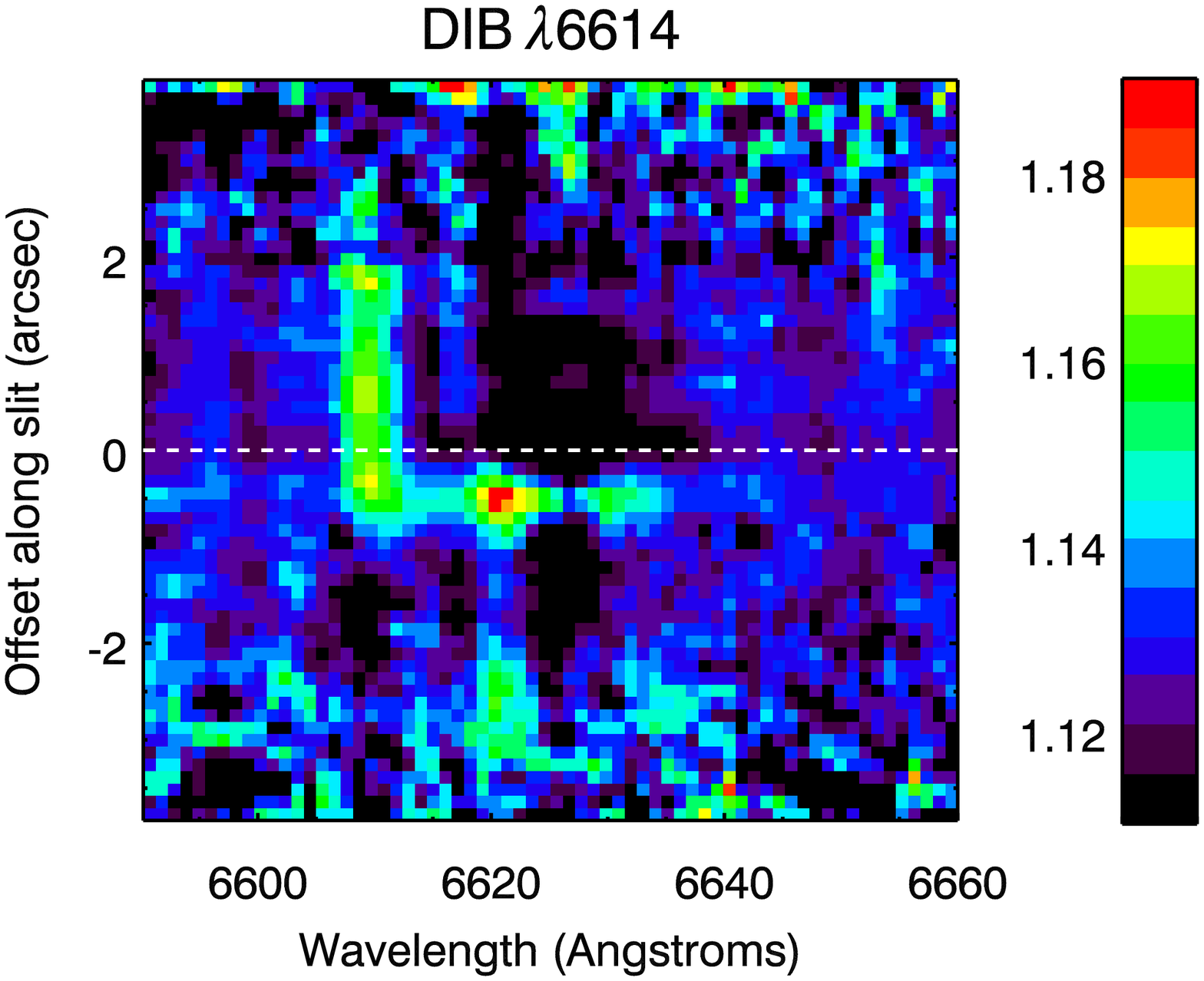}
  \caption{({\it Left:}) Normalised spectrogram of the unidentified
  blend from the synthetic long-slit spectrum, with slit aligned at
  33\degr east of north; ({\it right:}) the same spectrogram when
  divided through by the mean nebular spectrum. NE is the upper half
  of the spectrogram, the lower is SW. The dotted line shows the
  location of the intensity peak of the continuum, seen in similar,
  non-normalised images.}
  \label{fig:unls}
\end{figure}

\subsection{On the unidentified blend around the DIB at 6614\AA} \label{sec:un}
The bump of emission blueward of the DIB at 6614\AA\ is resolved in
\citet{Oudmaijer98} as four separate emission peaks (see his
Fig.\ 5). This is suggested by Oudmaijer to be a blend of Fe {\sc i},
although \citet{Humphreys02} suggests one line may be that of Sc {\sc
ii}. The feature is resolved as two peaks of similar strength in the
1998 spectrum of \citet{Chentsov99}, who suggest line identifications
of Sc {\sc ii} $\lambda$6605 and Ti {\sc ii} $\lambda$6607.

By taking a slice across the data-cube at a PA of 33\degr\ i.e.\
aligned with the symmetry identified from the H$\alpha$ \EW\ map, a
synthetic long-slit spectrogram of this feature was created (see
Fig.\,\ref{fig:unls}). The figure shows the feature on the blue side
of the DIB apparently behaving identically to the Ti {\sc ii} lines,
i.e.\ concentrated in the centre of the field. This is strong evidence
that this feature is in part due to Ti {\sc ii} emission, as suggested
by \citet{Chentsov99}. This feature {\it completely} vanishes to the
south, implying that all the unidentified, unresolved emission
components which make up this feature behave in the same way.


The spectrogram of this region also reveals the behaviour of the
unidentified feature on the red side of the DIB. This feature is
strongest $\sim$0.5\arcsec\ below the centre, disappears, then
reappears further south. The fact that it is apparently absent
elsewhere in the field may explain why this feature has never been
previously identified, as it is swamped by the rest of the emission
when integrated over a large region.

There is a possibility that must be discussed, that the features on
the red and blue side of the DIB are velocity-shifted components of
the {\it same} transition, formed in, for example, a bi-polar flow
viewed pole-on \citep[c.f.\ conclusions of][]{Oudmaijer98}. The velocity
separation of the two lines, which in this case is 450\kms, would be
centred on the systemic velocity of $\sim$75\kms\ \citep{Oudmaijer98},
and implies an outflow velocity of 225\kms. This means the rest-frame
wavelength of the transistion would be $6611.21 \pm 0.16$\AA,
neglecting any uncertainty in the systemic velocity.

As such behaviour is not observed in H$\alpha$ nor the singly-ionised
metallic lines of Fe and Ti, and as one may expect greatly different
physical conditions within the bi-polar flow to the rest of the wind,
one would expect this transition to be from a high ionization
species. The only such line in this narrow region is Ti {\sc iii}
$\lambda$6611.38. The presence of Ti {\sc iii} emission is not noted
anywhere in the high-resolution spectra of \citep{Oudmaijer98} and
\citet{Chentsov99}. This line-identification is therefore
unlikely. Further study of this feature with high SNR, high
spectral-resolution data is warranted to investigate the 'jet' nature
of this feature.

\subsection{The significance of wind-axisymmetry in IRC~+10420}
A wealth of evidence now exists for axisymmetry in the wind of
IRC~+10420 on several scales: from the outer reflection nebula seen
in {\it HST} images (PA$\sim$33\degr), high-resolution IR images on
scales of 1-2\arcsec\ \citep[$\sim$58\degr,][]{Humphreys97}, to the
scales of a few stellar radii probed by spectropolarimetry
($\sim$150\degr, Patel et al.\, {\it in prep}). The axes of symmetry
of these observations are all either aligned or perpendicular to
45$\pm$15\degr. 

The data presented in this paper are consistent with this story. The
quality of our data is such that we cannot resolve the changes in
line-profile when viewed from different angles, and so cannot draw any
conclusions as to the geometry of the wind (e.g.\ disk or bi-polar
ouflow) or density constrasts between equator and pole. However, the
\EW\ and velocity variations of the H$\alpha$ and Fe {\sc ii} lines
provide clear evidence for wind-axisymmetry, {\it independent of the
geometry of the reflection nebula}. The position-angle of this
axisymmetry is consistent with the long-axis of the nebula
($\sim$33\degr).

The axisymmetric wind of the star is strong evidence that rotation has
played a significant role in its evolution, whereby a
latitude-dependent effective gravity leads to very different mass-loss
behaviour between the equator and poles. This rotation could be due
either to the presence of a companion, or merely fast initial rotation
of the star. A rotating star which is losing mass is also losing
angular momentum, and hence the equatorial surface gravity should
steadily increase over time as mass is lost and the star spins
down. This may cause the mass-loss rate to decrease as a function of
age.

Studies of the star's dust emission by O96 and \citet{Blocker99} have
concluded that the mass-loss rate has decreased over time by factors
of 15 to 40; while \citet{Lipman00} also found a density gradient
consistent with a falling mass-loss rate. In addition, \citet{Smith04}
suggest that the gradual increase of the star's effective temperature
over the last $\sim$20 years or so
\citep[O96,][]{Oudmaijer98,Klochkova02} may be in fact due to a
further decrease in mass-loss. Here, the dense wind forms a {\it
pseudo-photosphere}, making the star appear cooler. As the mass-loss
rate falls and the wind density decreases, we are able to see deeper
into the wind, making the star's effective temperature appear to
increase.


Were the star's mass-loss rate to decrease further, the star would
continue to evolve blueward at constant bolometric luminosity. It is
then logical to connect IRC~+10420 with the B[e] supergiants (sgB[e]),
which occupy the same luminosity range on the HR diagram and whose
hybrid spectral characteristics are understood to result from
highly-axisymmetric outflows \citep[][]{Zickgraf86}. These stars are
often thought of as {\it preceding} the RSG phase
\citep[e.g.][]{Smith04}, a reasonable hypothesis given their large H
abundances (inferred from their strong Balmer-line emission),
presumably-high rotation-rates (inferred from their wind-axisymmetry)
and lack of circumstellar nebulae. Such stars could potentially shed
their angular momentum and spin-down upon passing through
the violent mass-losing RSG/YHG phases, leading the wind-geometry to
approach spherical symmetry. However, the data we present here shows that
IRC~+10420 still has significant residual wind-axisymmetry even after its
recent outbursts. Combined with its continued blueward evolution, it cannot
now be discounted that IRC~+10420 is evolving {\it toward} a sgB[e]-like
phase.

\section{Summary and conclusions} \label{sec:ifuconc}
This paper presents the first IFU spectroscopy of the nebula around
the transitional massive star IRC +10420. The key observational
results of this work can be summarized as follows:

\begin{itemize}
\item DIBs are strikingly uniform across the field, such that they
  cancel out perfectly when the total spectrum is normalised by the
  mean nebular spectrum. This is consistent with negligible
  circumstellar extinction towards the central star, as suggested by
  \citet{Jones93} and \citet{Oudmaijer98}.
\item The continuum subtracted, PSF-subtracted images of H$\alpha$ and
  Fe {\sc ii} both trace the extended morphology of the reflection
  nebula as shown in the $B$-band {\it HST} image. This supports the
  reflection-nebula paradigm of this object, as the star is unlikely
  to produce enough Lyman photons to ionize material out to the
  distances of the outer nebula.
\item The H$\alpha$ and Fe {\sc ii} lines both show a trend of
  decreasing equivalent width from SW to NE of about 25\%, following
  the long-axis of the slightly flattened reflection nebula. This is
  in contradiction to previous observations of this star by
  \citet{Humphreys02}. That the behaviour of the DIBs is so uniform is
  strong evidence that the \EW\ behaviour of H$\alpha$ and Fe {\sc ii}
  is real.
\item The velocity maps of the H$\alpha$ and Fe {\sc ii} lines appear
  very different, although both have an axi-symmetric geometry aligned
  roughly with the long-axis of the nebula. 
\item The images in the light of the Ti {\sc ii} emission {\it do not}
  trace the nebula morphology, in contrast to the corresponding
  H$\alpha$ and Fe {\sc ii} images. Indeed, the emission is apparently
  very close to being unresolved. This is inconsistent with the
  line-emission reflecting off the surrounding nebula. The line at
  6680\AA ~is broader than the one at 6718\AA, possibly due to a blend
  with He {\sc i} $\lambda$6678.
\end{itemize}

We conclude that the \EW\ behaviour of the H$\alpha$ and \Feii\ lines
is evidence for an axi-symmetric wind geometry in IRC +10420; and
that the centroid-shifts of these lines across the field are due to
complex line-profiles (due to e.g.\ anisotropic infall or B/R
variations) which appear different when viewed from different angles
by the surrounding nebula. That this behaviour was missed by
\citet{Humphreys02} is likely due to the unfortunate orientation of
their slit.

The \Tiii\ emission does not show the same behaviour as the H$\alpha$
and \Feii\ lines, in that the line-image does not trace the morphology
of the reflection nebula. It is speculated that this emission may
originate at the edge of the nebula, in an analogue of $\eta$ Car's
strontium filament, which also displays enhanced \Tiii\ emission. It is
noted that high-resolution spectra of IRC +10420 also show the other
trademark Sr filament lines of Fe {\sc i}, V {\sc ii} and Sr {\sc ii}.

Finally, we speculate that the present-day axisymmetry we observe in
IRC~+10420, combined with its continued blueward evolution at constant
bolometric luminosity, is evidence that the star is evolving {\it toward}
the B[e] supergiant phase.

\acknowledgments We thank the referee Nathan Smith for his suggestions
and comments which improved the paper, John Hillier for useful
discussions, and Willem-Jan de~Wit for a careful reading the
manuscript. We would also like to thank Chris Benn and Samantha Rix at
the WHT for their assistance during the observations. This work has
made use of the IDL software package, the GSFC IDL astronomy
libraries, and the HST data-archive. BD acknowledges
funding from PPARC.



 \bibliographystyle{apj}
 \bibliography{../../bibtex/biblio}

\end{document}